\documentclass[aps, pra,10pt, nofootinbib, amsmath, amssymb]{revtex4-2}

\usepackage{graphicx} 
\usepackage{amsmath}
\usepackage{amssymb}
\usepackage{physics}
\usepackage{float}
\usepackage{url}
\usepackage{hyperref}

\begin{document}

\title[Coherent Rydberg excitation using a pulsed fiber amplifier]{Coherent Rydberg excitation of single atoms using a pulsed fiber amplifier}
\author{Ying-Wen Zhang}
\affiliation{School of Mathematics and Physics, China University of Geosciences, Wuhan 430074, China}

\author{Yang Wang}
\affiliation{Division of Precision Measurement Physics, Innovation Academy for Precision Measurement Science and Technology, Chinese Academy of Sciences, Wuhan 430071, China}
\affiliation{School of Physical Sciences, University of Chinese Academy of Sciences, Beijing 100049, China}

\author{Chen-Long Xu}
\affiliation{Division of Precision Measurement Physics, Innovation Academy for Precision Measurement Science and Technology, Chinese Academy of Sciences, Wuhan 430071, China}
\affiliation{School of Physical Sciences, University of Chinese Academy of Sciences, Beijing 100049, China}

\author{Yi-Bo Wang}
\email{wangyb@apm.ac.cn}
\affiliation{Division of Precision Measurement Physics, Innovation Academy for Precision Measurement Science and Technology, Chinese Academy of Sciences, Wuhan 430071, China}
\affiliation{School of Physical Sciences, University of Chinese Academy of Sciences, Beijing 100049, China}

\author{Peng Xu}
\affiliation{Division of Precision Measurement Physics, Innovation Academy for Precision Measurement Science and Technology, Chinese Academy of Sciences, Wuhan 430071, China}
\affiliation{Wuhan Institute of Quantum Technology, Wuhan 430206, China}

\date{\today}

\begin{abstract}
In recent years, the growing scale of programmable neutral-atom arrays has led to an increasing demand for higher-power Rydberg excitation light. Although pulsed amplifiers deliver higher peak power than continuous-wave lasers, their use for efficient coherent Rydberg excitation of single atoms in arrays has been limited by challenges such as pulse distortion, synchronization with excitation sequences, and spectral linewidth broadening. Here, we address these issues using a fiber-based master-oscillator power-amplifier system. We demonstrate efficient coherent Rydberg excitation of single atoms in a rubidium atom array, achieving performance comparable to continuous-wave methods. This study provides a potentially new technical pathway toward future large-scale quantum simulation and computation with Rydberg atom arrays.
\end{abstract}



\maketitle

\section{Introduction}
Neutral-atom arrays have rapidly developed into a leading platform for realizing programmable quantum systems. In this approach, individual cold atoms are trapped in arrays of optical tweezers, and controllable interactions are introduced by exciting the atoms to Rydberg states. The system exhibits excellent qubit coherence \cite{Yang2016, Guo2020}, scalability \cite{Manetsch2025, Wang2026}, flexible qubit connectivity \cite{Bluvstein2022}, high-fidelity Rydberg excitation control \cite{Madjarov2020} and two-qubit logic gates \cite{Fu2022, Evered2023, Radnaev2025}. It has been successfully used to prepare complex quantum many-body states \cite{Scholl2021, Ebadi2021, Semeghini2021} and provides an experimental testbed for constructing scalable architectures toward universal fault-tolerant quantum computing \cite{Bluvstein2024, Bluvstein2026}.

To advance quantum simulation and computation, experimental systems are being progressively scaled up \cite{Preskill2018, Fan2018}, while placing higher demands on experimental control—particularly the need for higher-power Rydberg excitation light to address larger atom arrays and achieve faster Rabi frequencies \cite{Evered2023}. For example, in rubidium atoms, the 420\,nm + 1013\,nm two-photon excitation scheme is widely adopted, largely because readily available near-infrared continuous-wave lasers—such as Ti:sapphire and fiber lasers delivering tens to hundreds of watts—can efficiently drive the upper transition from the intermediate state to the Rydberg state, despite its small transition dipole moment. In neutral-atom arrays experiments, the sequence typically operates in a pulsed mode: the Rydberg excitation light is switched on only during short windows (typically on the order of microseconds) via components such as acousto-optic modulators (AOMs) and remains off otherwise. In such a pulsed configuration, the attainable peak power remains limited by the available continuous-wave laser power.

In contrast, pulsed lasers concentrate energy within extremely short time windows, delivering peak powers significantly higher ($>1$\,kW) than those of continuous-wave sources \cite{Zervas2014, Li2020, Yu2023}. This characteristic makes them a compelling alternative to continuous-wave lasers for supplying higher-power Rydberg excitation. Previous studies have utilized picosecond pulses for ultrafast Rydberg excitations \cite{Mahesh2025} and nanosecond pulses for electromagnetically induced transparency (EIT) spectroscopy \cite{Cai2025}. However, efficient coherent excitation of single atoms with microsecond pulsed lasers remains elusive, due to challenges such as severe pulse distortion \cite{Paschotta1997, Frantz1963}, synchronization with excitation sequences, and spectral linewidth broadening.

\begin{figure}[htbp]
 \centering
 \includegraphics[width=0.9\linewidth]{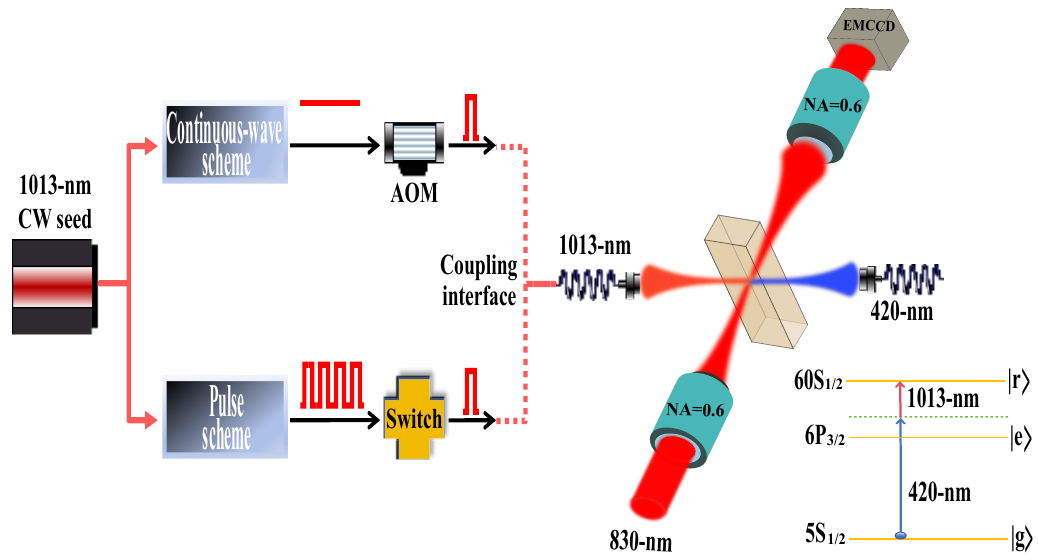}
 \caption{Schematic of the Rydberg excitation setup. An 830\,nm laser is used for single-atom trapping, and a two-photon excitation scheme at 420\,nm and 1013\,nm is employed to excite $^{87}$Rb atoms from the ground state to Rydberg state. The 1013\,nm light can be rapidly switched between a continuous-wave source and a pulsed MOPA via a pre-aligned fiber, enabling a side by side comparison of the two excitation methods.}
 \label{fig:1}
\end{figure}

In this paper, we implement a fiber-based master-oscillator power-amplifier (MOPA) scheme to achieve efficient coherent Rydberg excitation of single atoms in a rubidium atom array, as shown in Fig. \ref{fig:1}. We address the waveform distortion and the pulse-sequence synchronization. The demonstrated performance is comparable to that of continuous-wave methods, indicating that the MOPA architecture largely preserves the narrow linewidth of the seed laser during power amplification. This work thus provides a potentially new technical route toward scalable, high-fidelity Rydberg-based quantum control. The rest of this paper is organized as follows. Section \ref{sec:pre-shaping} describes the pre-shaping of the 1013\,nm seed-pulse waveform to suppress amplified pulse distortion. Section \ref{sec:excitation} presents the main experimental results, including a detailed implementation and direct comparison of the conventional continuous-wave excitation scheme and the pulsed MOPA scheme. Finally, Section \ref{sec:conclusion} provides a summary and discusses potential applications of the pulsed fiber amplifier scheme.

\section{Pulse pre-shaping compensates for waveform distortion during amplification}
\label{sec:pre-shaping}
In the experiment, a pulsed fiber amplifier module (Connet MFAP-Yb-1013-M-SF) with a peak power of 50\,W is employed to amplify the input 1013\,nm seed laser. This amplifier utilizes an ytterbium-doped fiber as the gain medium, with pump light establishing the initial population inversion between the upper and lower energy levels of the Yb$^{3+}$ ions. The seed laser, first modulated by an AOM to generate optical pulses with tunable pulse width and repetition rate, is then introduced into the gain medium. Within the fiber, photons from the seed pulse stimulate the excited Yb$^{3+}$ ions to return to the ground state, emitting additional photons through stimulated emission and thereby amplifying the pulse power.

\begin{figure}[htbp]
 \centering
 \includegraphics[width=0.9\linewidth]{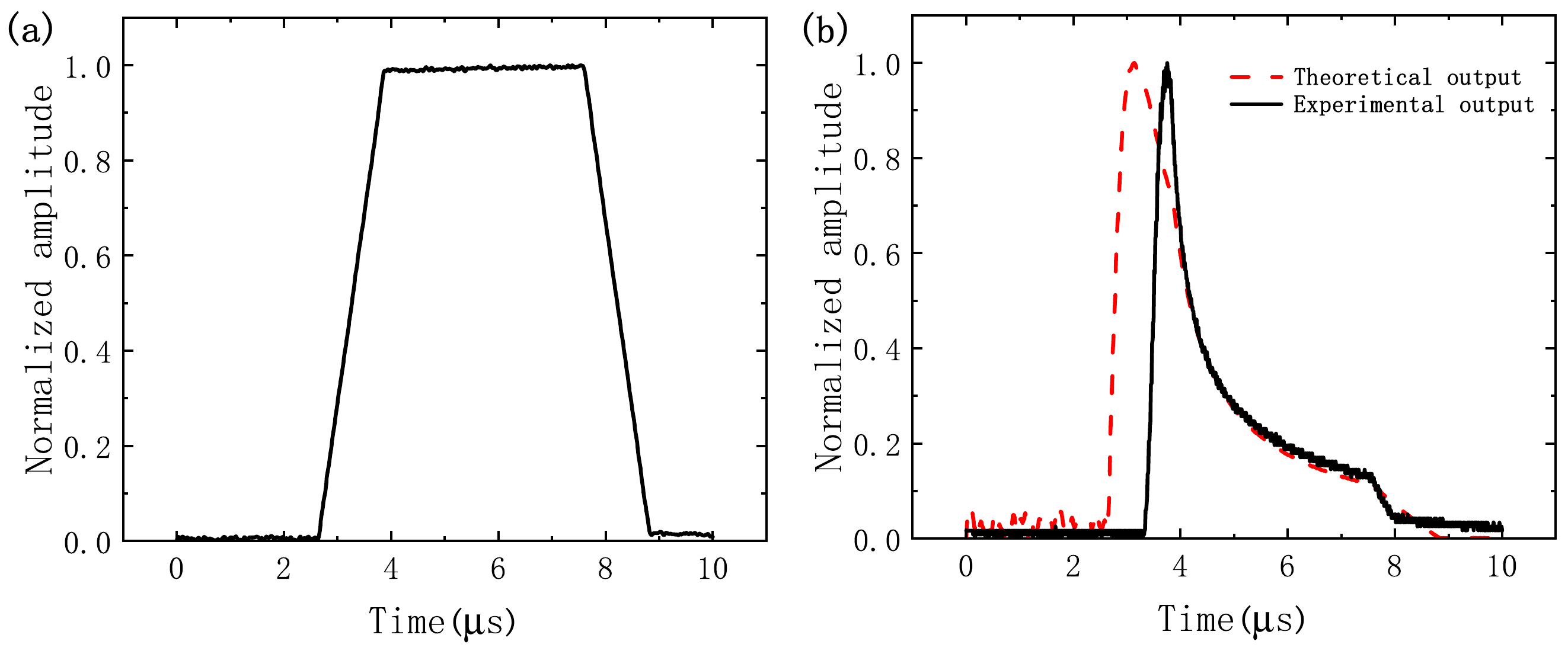}
 \caption{The shape of input and output pulses. (a) Input RF driving pulse. (b) Output pulse. The black and red curves denote the experimental and theoretical output waveforms, respectively. The amplifier module is operated at a pump power of 3\,W and a repetition rate of 2\,kHz.}
 \label{fig:2}
\end{figure}

During the amplification process, the leading edge of the pulse first enters the gain medium, where the population inversion has not yet been depleted, resulting in high gain and strong amplification. As the pulse propagates through the medium, the population inversion is progressively consumed. By the time the trailing edge arrives, the gain decreases, leading to weaker amplification \cite{Paschotta1997, Frantz1963}. This variation causes distortion in the output pulse waveform. Fig. \ref{fig:2} illustrates this phenomenon using a square input pulse as an example. The input waveform is shown in Fig. \ref{fig:2}(a), while the black curve in Fig. \ref{fig:2}(b) represents the experimental output waveform. It can be observed that the output waveform exhibits significant distortion due to the gain saturation effects.

The dynamic evolution between the photon density of the pulse and the initial population inversion during amplification can be quantitatively described by the Frantz–Nodvik equation \cite{Frantz1963}:
\begin{equation}
 n(z,t)=\frac{n_0(t - \frac{z}{c}) }{ 1 - [1 - e^{- \sigma \int_{0}^{z}\Delta_{0}(z')dz'} ]e^{-2\sigma c \int_{-\infty}^{t-\frac{z}{c}} n_0(t')dt'}} ,
 \label{eq:1}
\end{equation}
where $n$ denotes the photon number density, $\sigma$ is the stimulated emission cross-section, $c$ represents the speed of light in the medium, and $\Delta_{0}$ stands for the initial population inversion. Furthermore, we can derive the expression for the output pulse $I_{s,z}(T)$ after an input pulse $I_{s,0}(T)$ passes through a fiber of length $L$, which is given by:
\begin{equation}
 {I_{s,z}}\left ({T}\right ) = \frac{{I_{s,0}}\left ({T}\right )}{1-\left [{1-{G_0^{-1}}(z)}\right ]e^{{-{J_{sat}^{-1}}\int_{-\infty}^{T}{I_{s,0}}\left ({t'}\right )dt'}}}
 \label{eq:2}
\end{equation}
where $I$ is the optical intensity, $T=t-z/c$, $G_0(z)=\exp \left ( \sigma \int_{0}^{z} \Delta_0\left ({z'}\right )\mathrm{d}z'\right )$ is the small-signal gain, $J_{out}(T) = \int_{-\infty}^{T}{I_{s,0}}\left ({{t^{\mathrm{'}}}}\right )\mathrm{d}{t^{\mathrm{'}}}$ represents the output pulse fluence, and $J_{sat}$ is the saturation fluence, given by $h\nu/2\sigma$. At the onset of amplification, $J_{out} = 0$, denominator equals $1/G_0$, and the gain is $G_0$. As amplification proceeds, $J_{out}$ increases, causing the denominator to grow and the gain to gradually decrease. When $J_{out} \gg J_{sat}$, the denominator approaches unity, at which point the gain drops to 1, indicating that the gain is fully saturated and no further power amplification occurs.

Based on this model, the correspondence between the input and output waveforms can be quickly established. Taking the measured waveform in Fig. \ref{fig:2}(a) as the input, we adjust two undetermined parameters in Eq. (\ref{eq:2}), the initial population inversion and the stimulated emission cross-section, so that the theoretically calculated output waveform progressively approaches the measured output waveform. This process determines the values of these parameters. As indicated by the red curve in Fig. \ref{fig:2}(b), the theoretical waveform shows overall good agreement with the measured one. The main discrepancy is that the theoretical waveform is slightly broader than the measured one. This difference arises because the actual amplification process is accompanied by temporal pulse compression \cite{Luo2023}, an effect not accounted for in the theoretical calculation.

To correct the waveform distortion, we employ an input waveform pre-shaping scheme. Based on the established relationship between input and output waveforms, the Frantz–Nodvik equation is inversely solved to obtain\cite{Schimpf2008}:
\begin{equation}
 {I_{s,0}}\left ({t}\right )=\frac{{I_{s,z}}\left ({t}\right )}{1-\left [{1-G_0(z)}\right ]e^{{-{J_{sat}^{-1}}\int_{0}^{t}{I_{s,z}}\left ({t'}\right )dt'}}}
 \label{eq:3}
\end{equation}

\begin{figure}[htbp]
 \centering
 \includegraphics[width=0.9\linewidth]{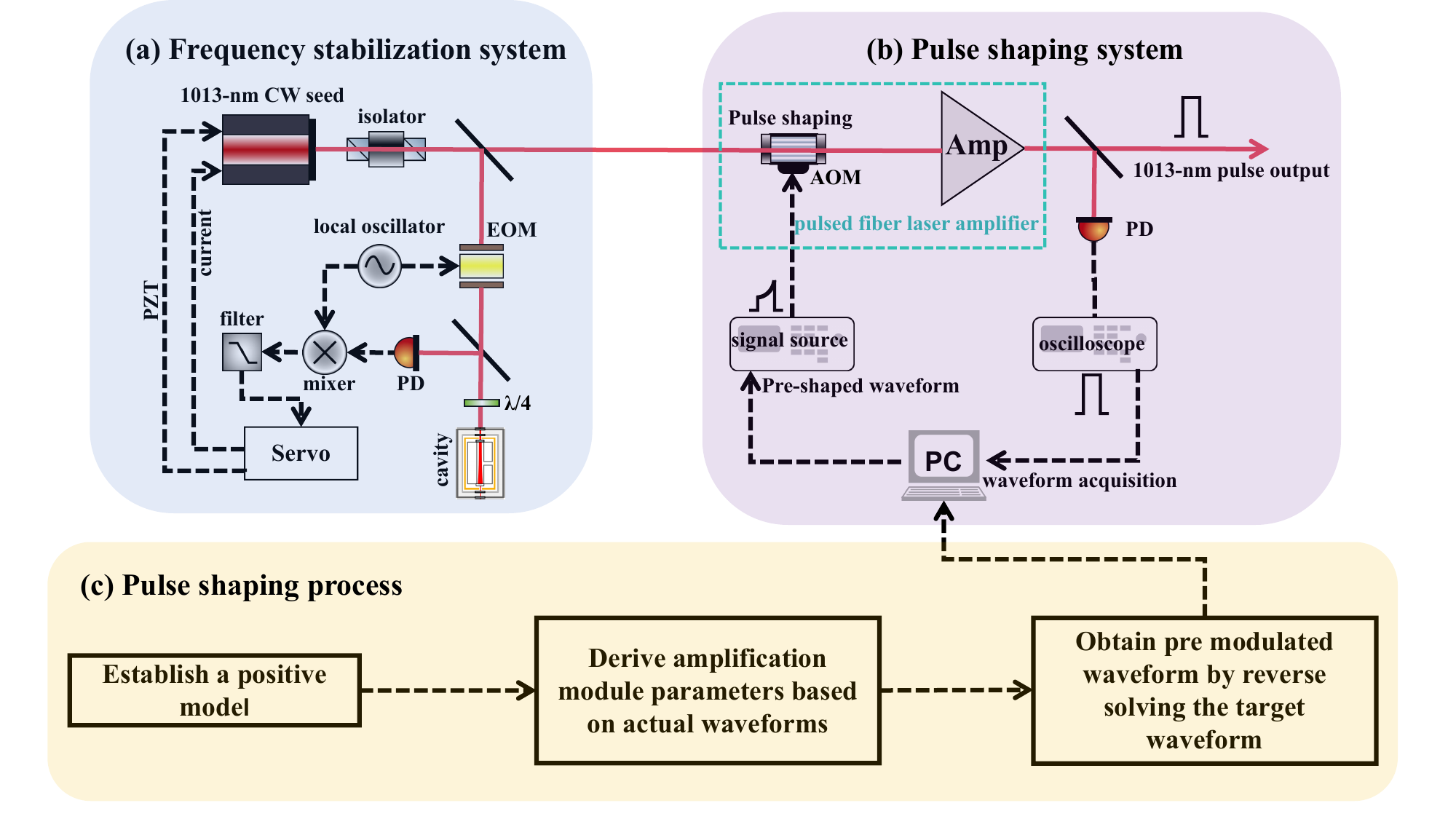}
 \caption{Waveform pre-shaping scheme. (a) Optical setup for Pound–Drever–Hall (PDH) locking of the laser to an ultra-stable cavity. (b) Schematic diagram of pulse shaping system. (c) The process of pulse shaping scheme.}
 \label{fig:3}
\end{figure}

Substituting the target output waveform into Eq. (\ref{eq:3}) yields the theoretical pre-shaped waveform. The basic workflow of waveform pre-shaping and the corresponding experimental setup are illustrated in Figs. \ref{fig:3}(b) and \ref{fig:3}(c). In practice, the theoretical pre-shaped waveform is loaded into a function generator to pre-modulate the seed light entering the pulsed fiber amplifier. The amplified optical pulse is detected with a high-speed photodetector; the resulting waveform is converted to an electrical signal and recorded by an oscilloscope. The amplified output waveform obtained directly from the theoretical pre-shaping still deviates from the target waveform because the experimentally applied pre-shaped waveform is the radio-frequency (RF) waveform that drives the internal AOM, not the optical pulse itself. Since the actual modulated optical pulse cannot be measured directly, iterative fine-tuning of the input waveform based on the measured output is required. To simplify this process, we start from the theoretical pre-shaped waveform, which can be well approximated by a combination of two piecewise functions: a linear rise followed by an exponential rise \cite{Vu2006}. This allows us to parameterize the waveform using only a few key parameters, rather than adjusting it point by point. By tuning these parameters, we conveniently adjust the input waveform. After each adjustment, the corresponding output waveform is recorded, normalized, and compared with the ideal target. This iteration is repeated until the desired output shape is achieved.

\begin{figure}[htbp]
 \centering
 \includegraphics[width=0.8\linewidth]{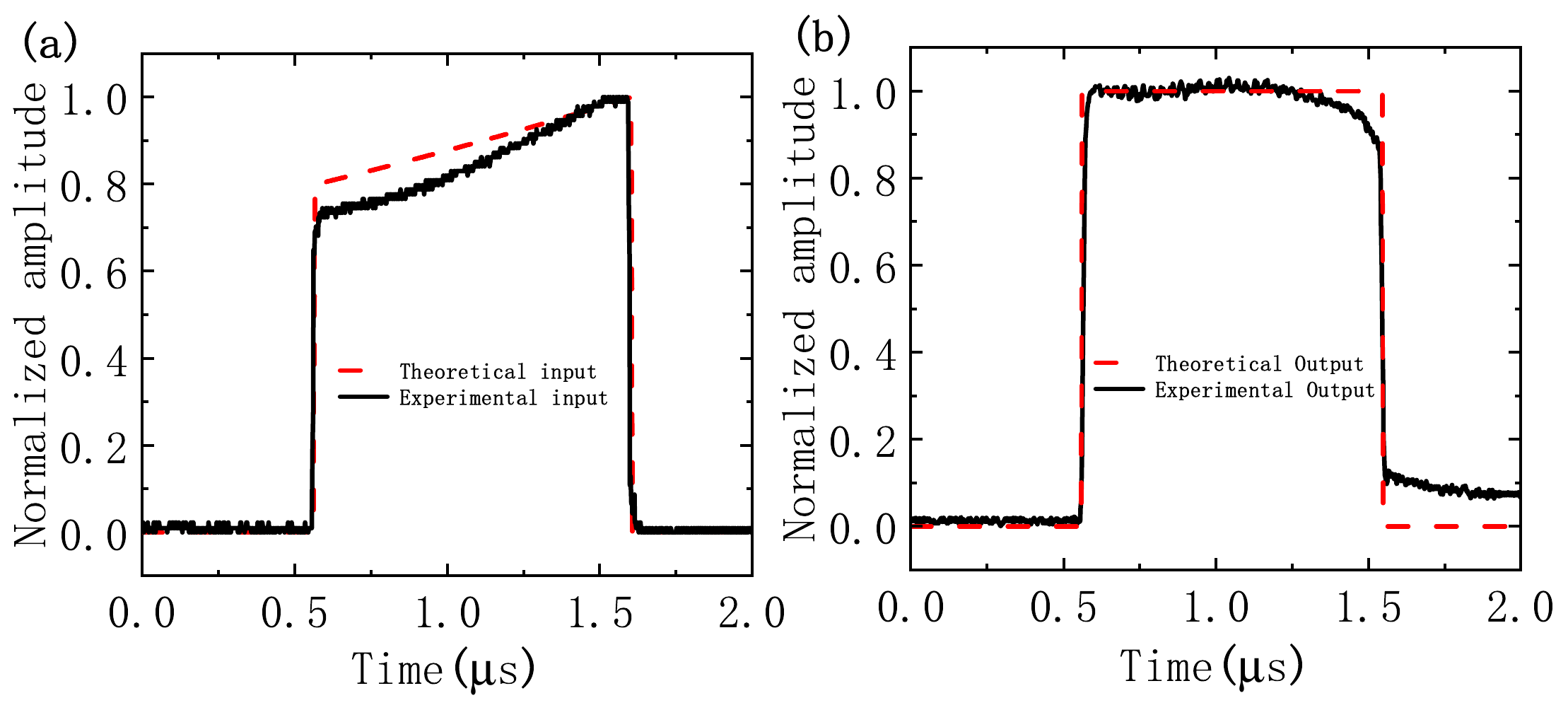}
 \caption{Waveform shaping. Comparison between theoretical calculated waveforms and experimental measured waveforms. (a) The comparison of input waveform. (b) The comparison of output waveform. The root mean square (RMS) value of deviation is 1\% in the interval between 0.6\,$\mu$s and 1.3\,$\mu$s. All red curves represent theoretical waveforms, while all black curves represent experimental waveforms. The amplifier module is operated at a pump power of 3\,W and a repetition rate of 2\,kHz.}
 \label{fig:4}
\end{figure}

The effectiveness of the pre-shaping scheme in suppressing output waveform distortion, illustrated here for a target 1\,$\mu$s square pulse, as shown in Fig. \ref{fig:4}. In Fig. \ref{fig:4}(a), the red curve shows the theoretically pre-shaped waveform, while the black curve corresponds to the experimentally implemented waveform after fine-tuning. The measured output (black curve) agrees well with the theoretical output (red curve), with a root mean square (RMS) deviation of 1\% over the interval between 0.6\,$\mu$s and 1.3\,$\mu$s, as shown in Fig. \ref{fig:4}(b). Compared to the unshaped input, the pre-shaped waveform effectively suppresses overall pulse distortion.

Although the output waveform distortion has been significantly reduced, it still deviates to some extent from an ideal square pulse. We attribute this discrepancy mainly to two factors. First, the waveform pre-shaping scheme we employed adjusts only a few key parameters to control the overall shape of the pre-modulated waveform, offering limited precision for finer details \cite{Vu2006, Xu2024}. Second, the pre-shaping approach is based on the Frantz–Nodvik equation, which inherently incorporates certain idealizations, such as neglecting nonlinear effects like stimulated Raman scattering(SRS) \cite{Stolen1972} and stimulated Brillouin scattering(SBS) \cite{Kobyakov2010}, which are unavoidable in practical systems.

\begin{figure}[H]
 \centering
 \includegraphics[width=0.8\linewidth]{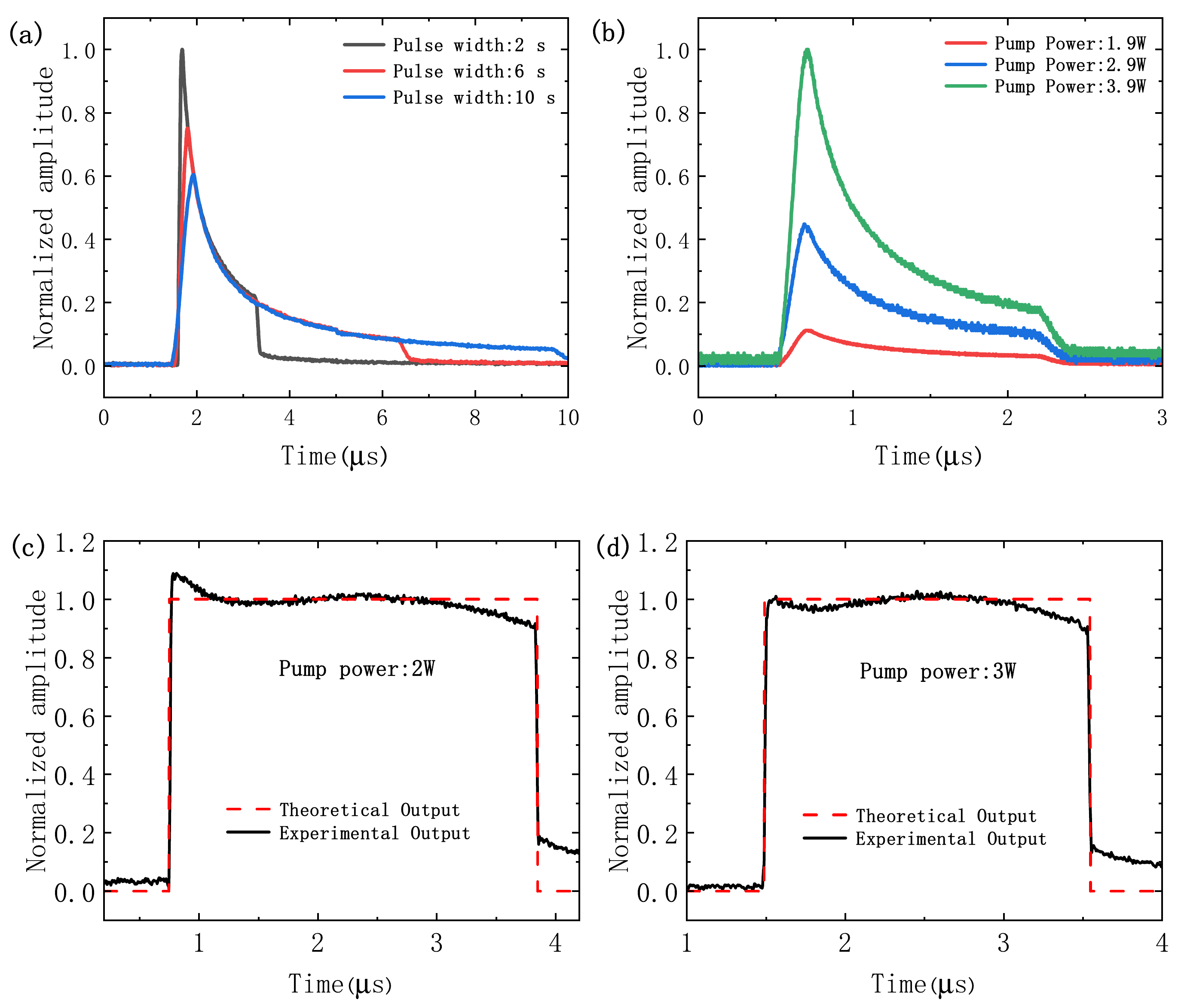}
 \caption{Output waveforms for square pulse inputs under different conditions. (a) Without pre-shaping, fixed pump power (1.9\,W) and varying pulse widths (2\,$\mu$s, 6\,$\mu$s, 10\,$\mu$s). (b) Without pre-shaping, fixed pulse width (10\,$\mu$s) and varying pump powers (1.9\,W, 2.9\,W, 3.9\,W). (c) With pre-shaping, output waveform achieving an RMS deviation of 1\% over the interval from 1.2\,$\mu$s to 3\,$\mu$s. (d) With pre-shaping, output waveform achieving an RMS deviation is 2\% over the interval from 1.8\,$\mu$s to 3.2\,$\mu$s.}
 \label{fig:5}
\end{figure}

In addition, we tested square pulses under different pulse widths and pump power conditions. The results show that when varying the pulse width at a fixed pump power, the output waveform distortion characteristics remain highly consistent because they lie on the same gain profile. Under a fixed pulse width, increasing the pump power raises the initial gain, extends the time required to enter the gain saturation regime, and consequently intensifies the waveform distortion. In both scenarios, applying the waveform pre-shaping method effectively suppresses the distortion and yields square pulse outputs that closely approximate the ideal shape, as shown in Figs. \ref{fig:5}.

In our experiment, the Rydberg excitation process is governed by the temporal overlap between the 420\,nm and 1013\,nm pulses. To minimize the influence of pulse shape imperfections, we position the 420\,nm pulse at the center of the 1013\,nm pulse, where the waveform is flat and stable. As a result, any deviations present in the leading or trailing edges of the 1013\,nm output waveform have negligible influence on the excitation process. The remaining small fluctuations in the flat-top region have only a minor effect. The current output waveform is therefore well suited for the subsequent experiments.

\section{Coherent Rydberg excitation of single atoms}
\label{sec:excitation}
We then apply the pre-shaped pulse to drive Rydberg excitations in an atom-array platform. First, a cloud of cold $^{87}$Rb atoms is prepared in a magneto-optical trap (MOT). Subsequently, an 830\,nm beam is phase-modulated by a spatial light modulator (SLM) and tightly focused through a high-numerical-aperture lens to form an optical tweezer array. The tweezer array overlaps spatially with the MOT and is capable of trapping up to $10 \times 20$ atoms. Individually loaded atoms are further cooled via polarization gradient cooling and adiabatic cooling, reaching a final temperature of approximately 7\,$\mu$K. Atoms are then optically pumped into the $|5S_{1/2}, F=1, m_F=0\rangle$ state and excited to a Rydberg state via a two-photon transition. Finally, fluorescence detection is used to measure the probability of atoms remaining in the ground state.

\begin{figure}[H]
 \centering
 \includegraphics[width=0.7\linewidth]{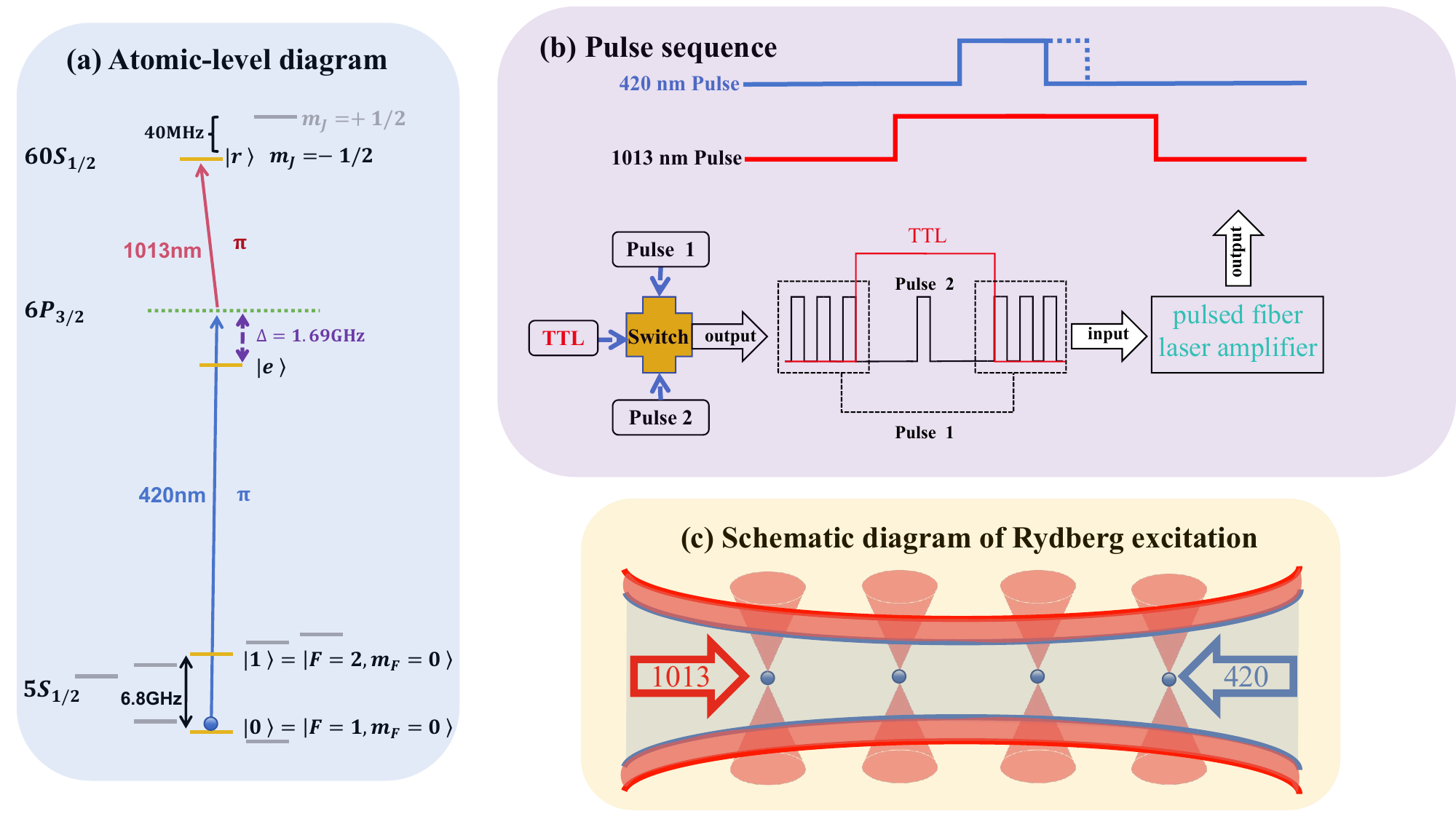}
 \caption{Rydberg excitation scheme. (a) Main energy level structure of $^{87}$Rb. (b) Pulse sequence. A single 1013\,nm pulse is selected from the pulse burst via a switch trigger for Rydberg excitation. The 420\,nm pulse is positioned in the middle of the 1013\,nm single pulse, and scanning the 420\,nm pulse time can yield Rabi oscillation. (c) Rydberg excitation geometry. Both the 420\,nm and 1013\,nm beams are aligned in a counter-propagating configuration to a row of four atoms separated by approximately 20\,$\mu$m.}
 \label{fig:6}
\end{figure}

The relevant energy-level structure of atom and the two-photon Rydberg excitation scheme, as shown in Fig. \ref{fig:6}(a). Atoms are excited from the $|0\rangle = |F=1, m_F=0\rangle$ state to the 60$S_{1/2}$ Rydberg state via a two-photon process. Both the 420\,nm and 1013\,nm excitation beams are $\pi$-polarized, with an intermediate-state detuning of 1.69\,GHz. The $\pi$-polarization of the Rydberg excitation light is adopted to facilitate the subsequent implementation of a magic-intensity optical trap scheme, ensuring the required magnetic field direction is perpendicular to the excitation beam \cite{Yang2016, Guo2020}. The 1013\,nm light originates from a homemade cat-eye laser; most of the output serves as seed light for the pulsed fiber amplifier, while a small portion is coupled into an ultra-stable cavity for frequency stabilization, resulting in a linewidth of a few hundred hertz (Fig. \ref{fig:3}(a)). The time sequence for Rydberg excitation and the method used to select a single 1013\,nm pulse, as shown in Fig. \ref{fig:6}(b). To maintain stable population inversion in the gain medium and prevent runaway accumulation of nonlinear effects, the pulsed fiber amplifier is operated in a burst mode \cite{Zervas2014}. For Rydberg excitation, a single pulse is extracted from the burst using a TTL-controlled switch. When the TTL signal is high, the amplifier outputs a single pulse (pulse 2) for excitation; when low, it continues to deliver the original pulse burst (pulse 1). The 420\,nm laser is generated by frequency-doubling an 840\,nm source, and its frequency is stabilized to the same ultra-stable cavity as the 1013\,nm laser. The Rydberg excitation geometry is shown in Fig. \ref{fig:6}(c). Two counter-propagating beams are aligned to a selected row of four atoms spaced about 20\,$\mu$m apart. At this spacing, interatomic interactions are negligible, allowing each atom to be excited independently. This configuration makes use of the uniformity of the excitation light while improving data acquisition efficiency.

To enable a direct comparison between the continuous-wave and pulsed schemes, the output of the pulsed amplifier is coupled into the pre-aligned 1013\,nm delivery fiber via a free-space optical path. This configuration allows rapid switching between the pulsed and continuous-wave excitation schemes as needed. Accounting for coupling losses in the free-space interface, approximately 50\% of the pulsed amplifier's power is available for excitation. Apart from the difference in the 1013 nm excitation sources themselves, all other experimental conditions are identical between the two schemes, as sketched in Fig. \ref{fig:1}.

The 1013\,nm Rydberg excitation pulse is generated using the pre-shaping scheme under conditions of 10\,$\mu$s pulse duration, 3.5\,W pump power, and a 2\,kHz repetition rate, corresponding to a peak power of approximately 5\,W from the pulsed fiber amplifier. Rabi oscillation is measured by scanning the duration of the 420\,nm pulse. To minimize decoherence due to power fluctuations the 420\,nm pulse is then positioned at the center of the flat region of the 1013\,nm pulse, as shown in Fig. \ref{fig:6}(b) \cite{Leseleuc2018}. The power fluctuation between amplified 1013\,nm Rydberg excitation pulses is approximately 2\%, while that for the 420\,nm excitation pulses is about 1\%. Since the Rabi frequency is proportional to the square root of the laser power, a few-percent shot-to-shot power fluctuation directly translates into a proportional fluctuation in the Rabi frequency. Experimentally, this manifests as an accelerated decoherence process and represents a main obstacle to achieving higher-fidelity quantum control in future implementations. In Section \ref{sec:conclusion}, we discuss how to further suppress power fluctuations in the pulsed scheme.

\begin{figure}[H]
 \centering
 \includegraphics[width=0.75\linewidth]{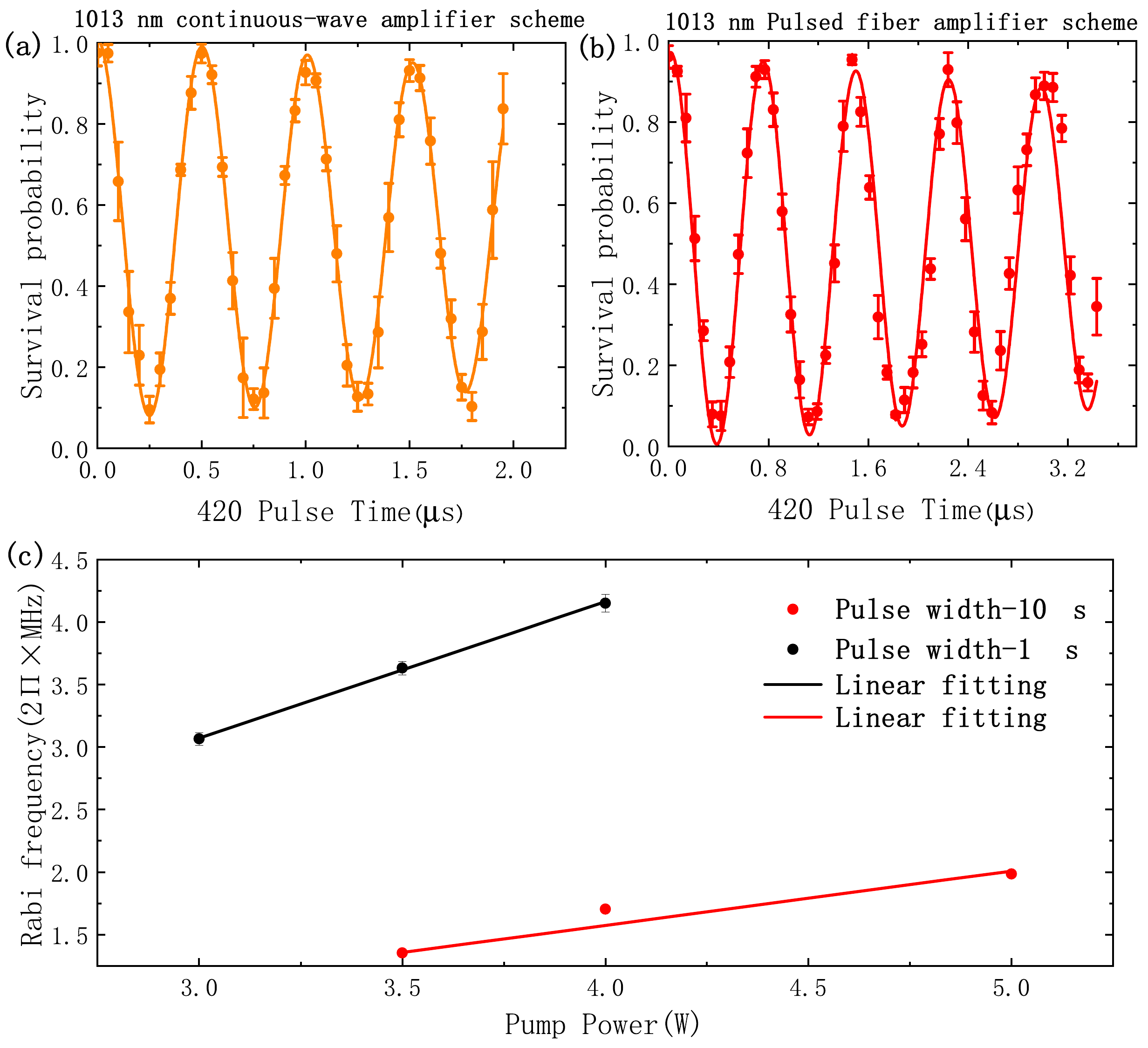}
 \caption{The Rabi oscillation between $|0\rangle$ and $|r\rangle$. (a) The Rabi oscillation fringes using the conventional continuous-wave amplifier scheme. Fitting the experimental data (orange solid points) yields a two-photon Rabi frequency of $2\pi \times (1.98 \pm 0.01)$\,MHz and the $1/e$ decay time is $11.14 \pm 3.62$\,$\mu$s. The shot-to-shot power fluctuation of the 1013\,nm excitation pulses is approximately 1\% for the CW scheme. (b) The Rabi oscillation fringes using the pulsed fiber amplifier scheme. Fitting the experimental data (red solid points) yields a two-photon Rabi frequency of $2\pi \times (1.35 \pm 0.01)$\,MHz and the $1/e$ decay time is $15.10 \pm 7.54$\,$\mu$s. The shot-to-shot power fluctuation is approximately 2\% for the pulsed scheme. All these data points are averaged from 60 or 150 experiment measurements, and the error bar represents the standard deviation of the mean. (c) The Rabi frequencies under different conditions of the pulsed fiber amplifier module.}
 \label{fig:7}
\end{figure}

We apply an external magnetic field to lift the degeneracy of the 60S state, splitting the $m_j$ sublevels by approximately 40\,MHz. By scanning the frequency of the 1013\,nm excitation laser, we locate the absorption peak corresponding to the transition from the ground state to the Rydberg state $|r\rangle = |60S, m_j=-1/2\rangle$ . With the 1013\,nm laser tuned to resonance, we then scan the duration of the 420\,nm pulse to drive coherent evolution between the ground and Rydberg states.

The Rabi oscillation between $|0\rangle$ and $|r\rangle$ obtained using the conventional continuous-wave scheme is shown in Fig. \ref{fig:7}(a). A fit to the experimental data (orange solid points) yields a two-photon Rabi frequency of $2\pi \times (1.98 \pm 0.01)$\,MHz, a $1/e$ lifetime of $11.14 \pm 3.62$\,$\mu$s, and a contrast of $0.94 \pm 0.03$. The corresponding oscillation measured with the pulsed fiber amplifier scheme is shown in Fig. \ref{fig:7}(b). Fitting the data (red solid points) yields a Rabi frequency of $2\pi \times (1.35 \pm 0.01)$\,MHz, a $1/e$ lifetime of $15.10 \pm 7.54$\,$\mu$s, and a contrast of $0.98 \pm 0.05$. Both schemes complete approximately 20 oscillation cycles within the $1/e$ lifetime, confirming their comparable coherence performance.

The pulsed fiber amplifier scheme enables the Rabi frequency to be increased directly by raising the pump power and shortening the pulse duration. To explore the performance limits of the amplifier module, we compare the achieved Rabi frequencies under different pump powers for two extreme pulse widths: 1\,$\mu$s and 10\,$\mu$s, and the result is shown in Fig. \ref{fig:7}(c). Under condition of 1\,$\mu$s pulse width, 2\,kHz repetition rate and pump power of 4\,W, the pulsed fiber amplifier reached its maximum output ($\sim 50$\,W peak power), yielding a two-photon Rabi frequency of $2\pi \times (4.25 \pm 0.06 )$\,MHz.

\begin{figure}[htbp]
 \centering
 \includegraphics[width=0.9\linewidth]{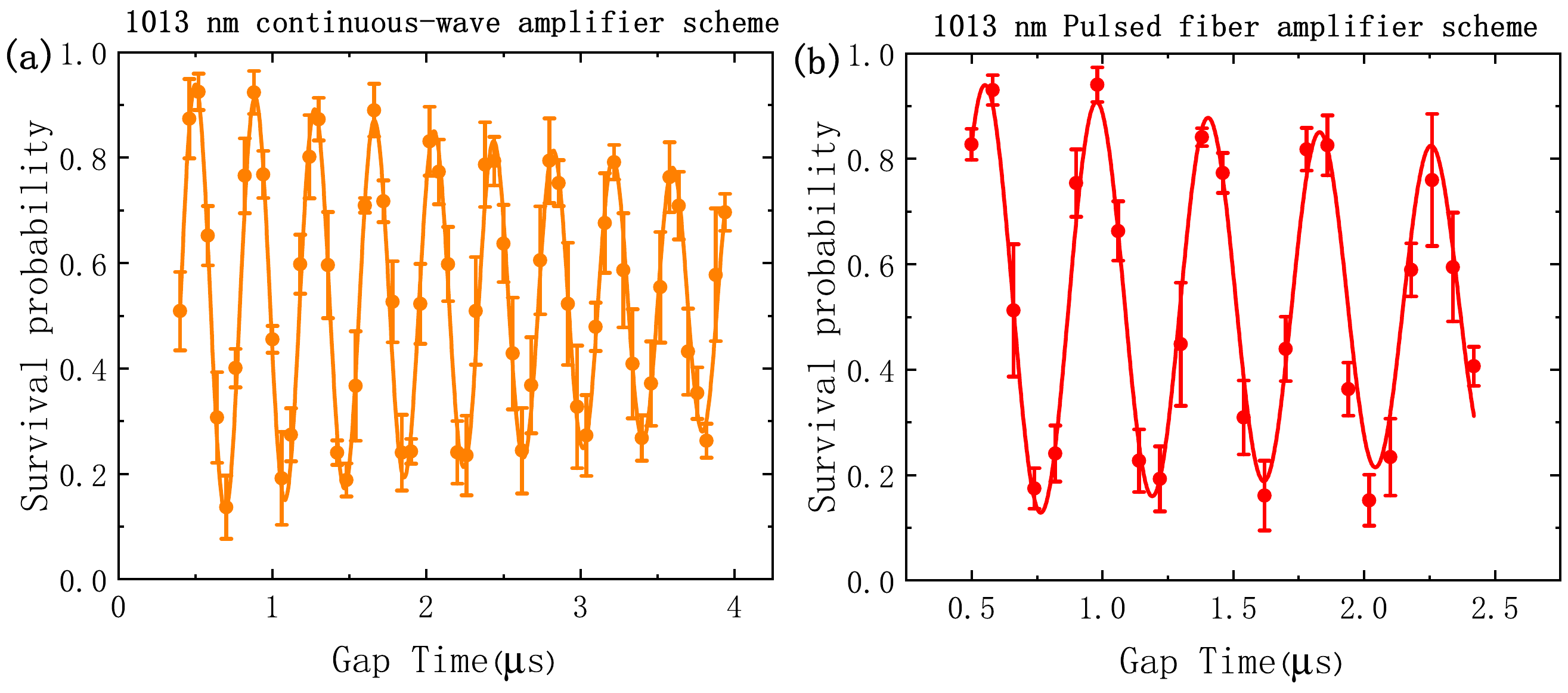}
 \caption{The Ramsey fringes between $|0\rangle$ and $|r\rangle$. (a) The Ramsey fringes using the continuous-wave amplifier scheme. The data points (orange) are fitted to exponentially decay sine curve (orange) with a $1/e$ decay time of $6.42 \pm 0.73$\,$\mu$s. The shot-to-shot power fluctuation of the 1013\,nm excitation pulses is approximately 1\% for the CW scheme. (b) The Ramsey fringes using the pulsed fiber amplifier scheme. The data points (red) are fitted to exponentially decay sine curve (red) with a $1/e$ decay time of $5.25 \pm 0.78 $\,$\mu$s. The shot-to-shot power fluctuation is approximately 2\% for the pulsed scheme. All these data points are averaged from 60 single experiment measurements, and the error bar represents the standard deviation of the mean.}
 \label{fig:8}
\end{figure}

To evaluate potential linewidth broadening introduced by the pulsed amplifier, we perform Ramsey interferometry between $|0\rangle$ and $|r\rangle$ using a $\pi/2$-gap-$\pi/2$ pulse sequence. The results for the conventional continuous-wave scheme and the pulsed fiber amplifier scheme are compared in Fig. \ref{fig:8}. For the pulsed fiber amplifier scheme, the fitted dephasing time extracted from the Ramsey fringes is $T_{2}^{\ast} = 5.25 \pm 0.78$\,$\mu$s (Fig. \ref{fig:8}(b)). At a temperature of $T=7$\,$\mu$K, the random shot-to-shot variation in atomic velocity results in a Gaussian-distributed Doppler detuning $\delta_{D} / (2\pi)$, corresponding to an expected dephasing time of 6.24\,$\mu$s. This indicates that the dephasing is dominated by thermal Doppler effects. The measured dephasing time for the continuous-wave scheme is $6.42 \pm 0.73$ \,$\mu$s (Fig.\ref{fig:8}(a)). The slight difference between the two can be attributed to two main factors. First, the shot-to-shot power fluctuation in the pulsed scheme is approximately twice that of the continuous-wave scheme; further stabilization of this fluctuation could help extend the dephasing time \cite{Liu2021}. Second, although the MOPA architecture largely preserves the narrow linewidth of the seed laser, the amplification process may introduce a slight increase in phase noise, which also contributes to the reduced dephasing time. Nevertheless, the difference in $T_{2}^{\ast}$ is modest, and the overall coherence performance of the pulsed scheme remains comparable to that of the continuous-wave scheme.

\section{Conclusion and Outlook}
\label{sec:conclusion}
This study demonstrates efficient coherent Rydberg excitation of single atoms using a fiber based MOPA system, achieving performance comparable to continuous-wave excitation. The MOPA architecture provides a promising route to deliver higher peak power while preserving spectral quality of the seed laser. The Rydberg Rabi frequency can be flexibly tuned through pump power and pulse width adjustments, potentially exceeding the coupling strength achievable with conventional continuous-wave lasers. This approach provides a scalable route to large-scale quantum simulation and computation based on Rydberg atom arrays. Compared with conventional continuous-wave schemes, the pulsed amplification approach offers significantly higher peak power. When combined with beam-shaping techniques, such as converting a Gaussian beam into a flat-top profile, this advantage enables either a larger illumination area at a given optical intensity or a higher intensity within a given area. Consequently, for a fixed target Rabi frequency, the pulsed scheme can efficiently drive a larger number of atoms, providing a scalable pathway toward larger neutral-atom arrays.

In the future, closed-loop feedback \cite{Oliveira2019, Yang2025} could further improve the amplified output waveform quality and enable arbitrary waveform generation. Such capabilities would be suitable for single off-resonant modulated pulse schemes \cite{Sun2023, Sun2020, Jandura2022} aimed at high-fidelity two-qubit gates. For those implementations, controlling the phase noise introduced during amplification—which has not been systematically examined here—remains an important task \cite{Vries2020, Sheng2021} and offers a compelling direction for further study. To further mitigate the decoherence effects induced by pulse power fluctuations, improvements in power stability can be achieved through several approaches, including enhancing the stability of the seed laser (e.g., active power stabilization and ensuring amplitude stability of the AOM driving signal), implementing current stabilization and temperature control for the pump source (e.g., using a low-noise current source combined with precision temperature control to reduce pump power fluctuations), and exploring active feedback control methods for power stabilization. Furthermore, the high-repetition-rate, short-pulse burst generated by the amplifier module are naturally suited for periodically driving quantum systems, enabling the study of novel nonequilibrium phenomena such as discrete time crystals \cite{Fan2020, Wang2025}.

\section*{Acknowledgments}
This research was supported by the National Key Research and Development Program of China under Grant No. 2021YFA1402001, the National Natural Science Foundation of China under Grants No. 12261131507, and No. 12074391, the Natural Science Foundation of Wuhan under Grant No. 2024040701010063, the Major Program (JD) of Hubei Province under Grant No. 2023BAA020.

\bibliographystyle{apsrev4-2}
\bibliography{refs}

\end{document}